\global\def\draftcontrol{0}

   \def\versionno{ bb effective action }

\catcode`\@=11

\expandafter\ifx\csname draftcontrol\endcsname\relax\global\def\draftcontrol{0}
\fi

{\count255=\time\divide\count255 by 60
\xdef\hourmin{\number\count255}
\multiply\count255 by-60\advance\count255 by\time
\xdef\hourmin{\hourmin:\ifnum\count255<10 0\fi\the\count255}}
\def\draftdate{\number\month/\number\day/\number\year\ \ \ \hourmin }

\newcommand\makepapertitle{\par
  \begingroup
    \renewcommand\thefootnote{\@fnsymbol\c@footnote}%
    \def\@makefnmark{\rlap{\@textsuperscript{\normalfont\@thefnmark}}}%
    \long\def\@makefntext##1{\parindent 1em\noindent
            \hb@xt@1.8em{%
                \hss\@textsuperscript{\normalfont\@thefnmark}}##1}%
     \newpage
     \global\@topnum\z@   
     \@makepapertitle
     \thispagestyle{empty}\@thanks
  \endgroup
  \setcounter{footnote}{0}%
  \global\let\thanks\relax
  \global\let\makepapertitle\relax
  \global\let\@makepapertitle\relax
  \global\let\@thanks\@empty
  \global\let\@author\@empty
  \global\let\@date\@empty
  \global\let\@title\@empty
  \global\let\title\relax
  \global\let\author\relax
  \global\let\date\relax
  \global\let\and\relax
  \def\version{\let\version\@version\@gobble}
}
\def\@makepapertitle{%
  \newpage
   \ifnum\draftcontrol=1 {}
   \version\versionno
   \vskip 3em%
   \else
   \hfill\hbox to 3cm {\parbox{4cm}{\@pubnum}\hss}%
   \vskip 3em%
   \fi
   \begin{center}%
   \let \footnote \thanks
     {\LARGE {\@title}}%
     \vskip 1.5em%
     {\normalsize
       \lineskip .5em%
       \begin{tabular}[t]{c}%
         \@author
       \end{tabular}\par}%
     \vskip 1.5em%
     {\@bstract}%
     \end{center}%
     \vskip 1.5em
     \@date%
   \par
}

\gdef\@pubnum{}
\def\pubnum#1{%
  \gdef\@pubnum{#1}}

\gdef\@bstract{}
\def\Abstract#1{%
  \gdef\@bstract{%
   \parbox{\textwidth-0pc}{%
   \centerline{\bf Abstract}\penalty1000%
\kern.2cm%
\noindent
\renewcommand\baselinestretch{1.0}%
{#1}}}
}

\def\ps@paper{\let\@mkboth\@gobbletwo%
     \ifnum\draftcontrol=1
    \def\@oddfoot{\hbox to \textwidth{\tiny \versionno \hfil\tiny\draftdate}%
    \hskip -\textwidth \hbox to \textwidth{\hfil\rm\thepage\hfil}}%
     \else\def\@oddfoot{\hbox to \textwidth{\hfil\rm\thepage\hfil}}
     \fi
     \let\@evenfoot\@oddfoot
}

\def\body{\clearpage
          \pagestyle{paper}
    }

\def\@version#1{\ifnum\draftcontrol=1
\typeout{}\typeout{#1}\typeout{}
\vskip3mm\centerline{\hbox{\fbox{\normalsize{\tt DRAFT -- #1 -- }
                   {\draftdate}}}}\vskip3mm
\fi}
\let\version\@version
\long\def\eqlabel#1{\ifnum\draftcontrol=1
                    \tag@false  
                    \tag*{(\theequation) \hbox to -0.2cm{\hspace{0cm}\small{#1}\hss}}
                    \refstepcounter{equation}
                    \edef\@currentlabel{\theequation}
                    \ltx@label{#1}          
                    \else
                    \label{#1}
                    \fi
                    }
\let\st@bibitem\@bibitem
\let\st@lbibitem\@lbibitem
\ifnum\draftcontrol=1
  \def\@bibitem#1{%
    \st@bibitem{#1}\a@@label{#1}\ignorespaces}
  \def\@lbibitem[#1]#2{%
    \st@lbibitem[#1]{#2}\a@@label{#2}\ignorespaces}
  \def\a@@label#1{%
    \gdef\a@lab{\smash{\normalfont\small#1}}
    \ifvmode
      \if@inlabel
        \global\setbox\@labels\hbox{%
          \llap{\a@lab\let\a@lab\relax
                \kern\@totalleftmargin\kern\marginparsep}%
          \box\@labels}%
      \fi
    \fi}
\fi

\documentclass[12pt,letterpaper]{article}

\usepackage{amsmath,amssymb,array,calc,epsfig,rotating,bm}
\usepackage[sort]{cite}
\usepackage{graphicx}
\usepackage{psfrag,verbatim}

\ifnum\draftcontrol=1
\fi

\renewcommand\baselinestretch{1.25}
\renewcommand\section{\@startsection {section}{1}{\z@}%
                                   {-3.5ex \@plus -1ex \@minus -.2ex}%
                                   {2.3ex \@plus.2ex}%
                                   {\normalfont\large\bfseries}}
\renewcommand\subsection{\@startsection{subsection}{2}{\z@}%
                                   {-3.25ex\@plus -1ex \@minus -.2ex}%
                                   {1.5ex \@plus .2ex}%
                                   {\normalfont\normalsize\bfseries}}
\renewcommand\subsubsection{\@startsection{subsubsection}{3}{\z@}%
                                   {-3.25ex\@plus -1ex \@minus -.2ex}%
                                   {1.5ex \@plus .2ex}%
                                   {\normalfont\normalsize\it}}
\renewcommand\paragraph{\@startsection{paragraph}{4}{\z@}%
                                   {-3.25ex\@plus -1ex \@minus -.2ex}%
                                   {1.5ex \@plus .2ex}%
                                   {\normalfont\normalsize\bf}}

\numberwithin{equation}{section}

\def\revise#1       {\raisebox{-0em}{\rule{3pt}{1em}}%
                     \marginpar{\raisebox{.5em}{\vrule width3pt\
                     \vrule width0pt height 0pt depth0.5em
                     \hbox to 0cm{\hspace{0cm}{%
                     \parbox[t]{4em}{\raggedright\footnotesize{#1}}}\hss}}}}

\newcommand\nxt[1]  {\\\fnxt#1}
\newcommand{\ie}{{\it i.e.,}\ }

\def\calz         {{\cal Z}}

\def\calb         {{\cal B}}

\def\calm         {{\cal M}}
\def\caln         {{\cal N}}
\def\calo         {{\cal O}}

\def\zet          {{\mathbb Z}}

\def\Re           {{\rm Re\hskip0.1em}}
\def\Im           {{\rm Im\hskip0.1em}}

\def\sqr#1#2{{\vcenter{\vbox{\hrule height.#2pt
 \hbox{\vrule width.#2pt height#1pt \kern#1pt
 \vrule width.#2pt}\hrule height.#2pt}}}}

\newcommand{\ft}[2]{{\textstyle{\frac{#1}{#2}}}}

\newcommand{\kk}{\mathfrak{q}}

\def\w{\omega}

\def\dd{\delta}
\def\e{\epsilon}

\def\aa1{\phi}
\def\cc1{\psi}

\def\k{\kappa}

\def\l{\lambda}

\def\Om{\Omega}
\def\om{\Omega}

\def\tf{\tilde{f}}

\def\uE{\underline{E}}
\def\t{\tau}

\catcode`\@=12

\begin{document}


\title{\bf Effective Action of the Baryonic Branch in String Theory Flux Throats }
\pubnum{UWO-TH-14/4}

\date{July 18, 2014}

\author{
Alex Buchel\\[0.4cm]
\it Department of Applied Mathematics\\
\it University of Western Ontario\\
\it London, Ontario N6A 5B7, Canada\\[0.2cm]
\it Perimeter Institute for Theoretical Physics\\
\it Waterloo, Ontario N2J 2W9, Canada\\[0.2cm]
 }

\Abstract{We discuss consistent truncations of type IIB supergravity on resolved
warped deformed conifolds with fluxes.  These actions represent the
gravitational duals to the baryonic branch deformation of the
Klebanov-Strassler cascading gauge theory. As an application, we
demonstrate that the baryonic branch is lifted in cascading gauge
theory plasma.
}

\makepapertitle

\body

\version\versionno
\tableofcontents

\section{Introduction and Summary}
A conifold $Y_6$ is a simplest non-compact Calabi-Yau three-fold \cite{Candelas:1989js}.
It is a cone over a homogeneous five dimensional Einstein manifold $T^{1,1}=(SU(2)\times SU(2))/U(1)$, with the 
$U(1)$  being a diagonal subgroup of the maximal torus of $SU(2)\times SU(2)$.
When a large number $N\gg 1$ of D3-branes are placed at its tip, for large 
't Hooft coupling $g_s N\gg 1$ their backreaction {\it warps} the conifold:
\begin{equation}
R^{3,1}\times Y_6 \qquad \to \qquad AdS_5\times T^{1,1}\,.
\end{equation}
Along with $N$-units of 5-form flux through $T^{1,1}$, the resulting geometry 
is a consistent background of type IIB string theory, holographically dual to $\caln=1$ 
four-dimensional superconformal $SU(N)\times SU(N)$ gauge theory \cite{Klebanov:1998hh}.
The warped conifold can be {\it deformed} (without further breaking the supersymmetry)
by wrapping $M\gg 1$ D5-branes over the two-cycle of $T^{1,1}$. In this case the 
supergravity background realizes the holographic dual to non-conformal $\caln=1$ supersymmetric 
$SU(N+M)\times SU(N)$ cascading gauge theory \cite{Klebanov:2000hb} (KS). One the geometry side, 
the $SU(2)\times SU(2)\times U(1)$ global symmetry of $T^{1,1}$ is broken to $SU(2)\times SU(2)\times \zet_2$.
The conifold deformation parameter breaking $U(1)\to \zet_2$ represents the spontaneous chiral symmetry 
breaking in the confining vacuum of cascading gauge theory. 
The vacuum structure of  $\caln=1$ cascading gauge theories was studied in  
\cite{Dymarsky:2005xt}. Precisely when $N$ is an integer multiple of $M$, there is a baryonic 
branch of confining vacua. In fact, the KS vacuum (without mobile D3-branes) corresponds to a special $\zet_2$ symmetric 
point on this branch.  A generic point on the baryonic branch breaks $\zet_2$. The supergravity dual to the
baryonic branch of cascading gauge theory was constructed in \cite{Butti:2004pk} (BGMPZ): moving 
away from the $\zet_2$ symmetric solution corresponds to a {\it resolution} of the KS warped deformed 
conifold. 

The type IIB supergravity backgrounds constructed in \cite{Klebanov:2000hb} and \cite{Butti:2004pk}
are supersymmetric, and thus are not suitable to address nonsupersymmetric questions in cascading gauge theory. 
Likewise, given the prominent role the KS warped throat geometries play in constructing de-Sitter vacua
in string theory \cite{Kachru:2003aw}, one needs to understand generic nonsupersymmetric 
deformations of BGMPZ {\it resolved warped deformed} conifolds. The first step in this direction was 
taken in \cite{Aharony:2005zr}, where a five dimensional effective action describing the 
$SU(2)\times SU(2)\times U(1)$ invariant sector of the warped conifold was constructed. This action 
includes five dimensional metric coupled to four bulk scalar fields. It was used to 
prove the renomalizability of cascading gauge theory \cite{Aharony:2005zr}, and  
detailed studies of thermodynamics and hydrodynamics of chirally symmetric 
phase of cascading gauge theory plasma  \cite{Buchel:2005cv,Aharony:2007vg,Buchel:2009bh}.  
In \cite{Aharony:2007vg} it was shown that cascading gauge theory undergoes the first order 
confinement-deconfinement phase transition at a certain critical temperature $T_c$. Furthermore,
there is a critical point at $T_u=0.8749(0) T_c$ where the  chirally symmetric phase 
becomes perturbatively unstable towards condensation of hydrodynamic (sound) modes \cite{Buchel:2009bh}. 
To understand chiral symmetry breaking in cascading gauge theory plasma, in \cite{Buchel:2010wp}
we derived  effective action corresponding to $SU(2)\times SU(2)\times \zet_2$ invariant 
sector of the warped deformed conifold --- here, three additional scalar fields are included compare 
to \cite{Aharony:2005zr}. This effective 
action\footnote{Additional applications were considered in \cite{Buchel:2011cc,Buchel:2013dla}.} 
was used to establish that chiral symmetry breaking 
fluctuations in cascading gauge theory plasma  become tachyonic at $T_{\chi {\rm SB}} =0.882503(0) T_c $; 
as a result, both confinement and the chiral symmetry breaking in cascading plasma occur simultaneously 
via the first-order phase transition at $T_c$. 

Comparing to the warped deformed conifold consistent truncation  
\cite{Buchel:2010wp}, the  BGMPZ supersymmetric holographic renormalization group (RG) flow contains 
two additional scalar fields (a mode dual to a dimension two operator and a mode mode dual to a dimension four 
operator of the boundary cascading gauge theory). It is straightforward to perform Kaluza-Klein reduction 
of this enlarged gravity-scalar sector and produce a five-dimensional truncation
of the resolved warped deformed conifold \cite{Berg:2005pd}\footnote{See also \cite{Buchel:2014hja}.}. 
Unfortunately,  this action is not a consistent truncation away from the origin of the baryonic branch
\cite{Berg:2005pd}\footnote{I would like 
to thank Davide Cassani and Anton Faedo
for bringing reference \cite{Berg:2005pd} to my attention, and pointing out the 
inconsistency of the truncation \cite{Buchel:2014hja}.}; 
at the origin of the baryonic branch the truncation is consistent and is identical to \cite{Buchel:2010wp}.  

The fully consistent $SU(2)\times SU(2)$ truncation of type IIB supergravity on resolved warped deformed
conifold was constructed in  \cite{Cassani:2010na}\footnote{Related discussion appeared in 
\cite{Bena:2010pr}. We will not attempt to verify \cite{Bena:2010pr} and relate it to earlier work,
partly because the authors did not present the Chern-Simons part of the action in full generality. } (CF).
In this paper we reproduce the derivation of the effective action \cite{Cassani:2010na}, and 
point further consistent truncation to  
effective action \cite{Buchel:2010wp}. We further discuss linearized fluctuations of 
CF effective action about $SU(2)\times SU(2)\times U(1)$ symmetric warped conifold with fluxes 
consistent truncations of \cite{Aharony:2005zr}. We recover consistent truncation of chiral symmetry breaking 
sector in cascading gauge theory plasma \cite{Buchel:2010wp}. Lastly, we present linearized 
effective action describing baryonic branch deformation about  $SU(2)\times SU(2)\times U(1)$ symmetric states 
of cascading gauge theory plasma.  We show that unlike $\zet_2$-invariant chiral symmetry breaking fluctuations, 
$\zet_2$-non-invariant baryonic branch fluctuations remain massive up to $T_u$ in cascading 
gauge theory plasma, \ie the baryonic branch is lifted by the finite temperature effects.

\section{Effective action}\label{section2}
In this section, following  \cite{Cassani:2010na} and   
\cite{Cassani:2010uw}\footnote{Related work appeared also in 
\cite{Gauntlett:2010vu}.},
we reproduce  the  derivation of CF effective action of the 
resolved warped deformed conifold with fluxes. The offshoot is that the effective action derived in 
\cite{Cassani:2010na} is correct; moreover, we did not find any typos in the 
presentation.

We will work in the gravitational approximation to type IIB string theory,
using the type IIB supergravity action. This action takes the form (in 
the Einstein frame)
\begin{equation}
\begin{split}
S_{10}=&\frac{1}{2\k_{10}^2}\int_{\calm_{10}}\ \biggl(
R_{10}\wedge \star_{10} 1 -\frac 12 d\phi\wedge \star_{10} d\phi
-\frac 12 e^{-\phi} H_3\wedge\star_{10} H_3  -\frac 12 
e^{\phi} F_3\wedge\star_{10} F_3\\
&\qquad \qquad -\frac 12 e^{2\phi} F_1\wedge \star_{10} F_1
 -\frac 14 F_5\wedge \star_{10} F_5\biggr)\\
&-\frac{1}{8\k_{10}^2}\int_{\calm_{10}} \left(B_2\wedge d(C_2)
-C_2 \wedge d(B_2)\right)\wedge d(C_4),
\end{split}
\eqlabel{10action}
\end{equation} 
where $\calm_{10}$ is the ten dimensional bulk space-time, $\k_{10}$ is
the ten dimensional gravitational constant. The form-field strengths,
determined  from the potentials $\{C_0$ $,B_2$ $,C_2$ $,C_4\}$, satisfy the 
Bianchi identities:
\begin{equation}
 d(F_1)=0\,,\qquad d(H_3)=0\,,\qquad d(F_3)=H_3\wedge F_1\,,\qquad d(F_5)=H_3\wedge F_3\,.
\eqlabel{flbianchi}
\end{equation}
The equations of motion following from the 
action \eqref{10action} have to be supplemented by
the self-duality condition
\begin{equation}
\star_{10} F_5=F_5\,.
\eqlabel{5self}
\end{equation}
It is important to remember that the self-duality 
condition \eqref{5self} 
can not be imposed at the level of the action,
as this would lead to wrong equations of motion.

Appendix \ref{apA} contains our conventions regarding differential forms.

\subsection{Left-invariant forms on the $T^{1,1}$ coset}

We use explicit parametrization of the coset $T^{1,1}= (SU(2)\times SU(2))/ U(1)$ 
in terms of angular coordinates $\{\theta_1,\phi_1,\theta_2,\phi_2,\psi\}$ with 
ranges $0\le \theta_{1,2}<\pi$, $0\le \phi_{1,2}<2\pi$, and $0\le \psi <4\pi$. As in 
\cite{Cassani:2010na} we choose the coframe 1-forms as 
\begin{equation}
\begin{split}
&e^1=-\sin\theta_1\ d(\phi_1)\,,\qquad e^2=d(\theta_1)\,,\\
&e^3=\cos\psi\ \sin\theta_2\ d(\phi_2)-\sin\psi\ d(\theta_2)\,,\\
&e^4=\sin\psi\ \sin\theta_2\ d(\phi_2)+\cos\psi\ d(\theta_2)\,,\\
&e^5=d(\psi)+\cos\theta_1\ d(\phi_1)+\cos\theta_2 d(\phi_2)\,. 
\end{split}
\eqlabel{coframe}
\end{equation} 
All left-invariant 1- and 2-forms on the coset are given by \cite{Cassani:2010na}:
\begin{equation}
\begin{split}
&\eta=-\frac 13 e^5\,,\qquad \Omega=\frac 16 (e^1+i e^2)\wedge (e^3-i e^4)\,,\\
&J=\frac 16 (e^1\wedge e^2-e^3\wedge ^4)\,,\qquad \Phi=\frac{1}{6}(e^1\wedge e^2+e^3\wedge e^4)\,. 
\end{split}
\eqlabel{lif}
\end{equation}

\subsection{Metric ansatz and its dimensional reduction}
We take the ten-dimensional spacetime $\calm_{10}$ to be a direct warped product $\calm_5\times T^{1,1}$.
The most general $SU(2)\times SU(2)$ invariant metric on $\calm_{10}$ is parameterized by five 0-forms 
$\{u,v,\t,\w,\theta\}$, and a single 1-form $A$ on $\calm_5$,  \cite{Cassani:2010na}:
\begin{equation}
\begin{split}
&ds_{\calm_{10}}^2=\sum_I\uE^I \uE^I\,,\qquad ds_{\calm_5}^2=\sum_i E^i E^i\,,\\
&\uE^I=e^{-\ft 43 u -\ft 13 v} E^i\,,\qquad {\rm for}\qquad I=i=1,\cdots5\,, \\
&\uE^6=\frac{1}{\sqrt{6\cosh \t}}\ e^{u+w}\ e^1\,,\qquad \uE^7=\frac{1}{\sqrt{6\cosh \t}}\ e^{u+w}\ e^2\,,\\
&\uE^8=\sqrt{\frac{\cosh\t}{6}}\ e^{u-w}\ \left(\ e^3+\tanh\t\ e^{2\w}\ \Re\left(e^{i \theta}(e^1+i e^2)\right)\ \right)\,,\\
&\uE^9=\sqrt{\frac{\cosh\t}{6}}\ e^{u-w}\ \left(\ e^4+\tanh\t\ e^{2\w}\ \Im\left(e^{i \theta}(e^1+i e^2)\right)\ \right)\,,\\
&\uE^{10}=e^v (\eta+A)\,.
\end{split}
\eqlabel{10frame}
\end{equation} 
Given \eqref{10frame}, it is straightforward (albeit tedious) to reduce ten-dimensional Einstein-Hilbert term in 
\eqref{10action}. We find 
\begin{equation}
\begin{split}
&\frac{1}{2\k_{10}^2}\int_{\calm_{10}} (R\star_{10} 1) = \frac{1}{2\k_5^2}\int_{\calm_5}\biggl[
R -\frac 12 e^{\ft 83 u +\ft 83 v}\ (dA)^2+e^{-\ft 83 u -\ft 23 v}\ R_{T^{1,1}}\\
&-\frac{28}{3} du^2-\frac 43 dv^2-\frac 83 dudv -d\t^2-4\cosh^2 \t\ d\w^2
-\sinh^2\t\ (d\theta-3 A)^2
\biggr]\star 1\,,
\end{split}
\eqlabel{ehred}
\end{equation}
where 
\begin{equation}
\begin{split}
R_{T^{1,1}}=&4 e^{-4 u +2 v}\ \left[ \sinh^2\t -\cosh^2\t \ \cosh(4\w)\right]
\\&+24 e^{-2u}\cosh\t\ \cosh(2\w)-9 e^{-2 v}\ \sinh^2 \t\,,
\end{split}
\eqlabel{rt11}
\end{equation}
and 
\begin{equation}
\k_5^2=\frac{\k_{10}^2}{V_Y}\,,\qquad V_Y=-\frac 12 \int_{T^{1,1}} J\wedge J\wedge \eta\,,
\eqlabel{defk5}
\end{equation}
with $V_Y$ being the volume of unit size $T^{1,1}$. 

$SU(2)\times SU(2)$ symmetry requires that both the dilaton $\phi$ and the axion $C_0$ are 
0-forms on $\calm_5$. Their reduction on $T^{1,1}$ is trivial:
\begin{equation}
\begin{split}
\frac{1}{2\k_{10}^2}\int_{\calm_{10}}\left(-\frac 12(d\phi)^2-\frac 12 e^{2\phi} F_1^2\right)\star_{10} 1
=-\frac{1}{2\k_5^2}\int_{\calm_5}\ \left[\frac 12 d\phi^2+\frac 12 e^{2\phi} dC_0^2\right]\star 1\,.
\end{split}
\eqlabel{axidilaton}
\end{equation}

\subsection{3-forms ansatz and their dimensional reduction}

Most general $SU(2)\times SU(2)$ symmetric ansatz of NSNS 3-form field strength $H_3$ 
(solving the Bianchi identity \eqref{flbianchi}) is 
parameterized by a 2-form $b_2$, a one form $b_1$, two real 0-forms $b^J$ and $b^\Phi$, a 
complex 0-form $b^\Omega$ on $\calm_5$ and a constant $p$, 
 \cite{Cassani:2010na}:
\begin{equation}
\begin{split}
H_3=&p\ \Phi\wedge \eta+ d(B_2)\,,\\
B_2=&b_2+ b_1\wedge (\eta+A)+b^J J+\Re(b^\Om\Om)+b^\Phi\Phi\,.
\end{split}
\eqlabel{defb2h3}
\end{equation}
The field strength $H_3$ can be decomposed in a basis of left-invariant forms on $T^{1,1}$ \eqref{lif}:
\begin{equation}
\begin{split}
H_3=&h_3+h_2\wedge (\eta+A)+h_1^J\wedge J+\Re\left[h_1^\Om\wedge \Om+h_0^\Om\ \Om\wedge (\eta+A)\right]\\
&+h_1^\Phi\wedge \Phi+p\ \Phi\wedge (\eta+A)\,,
\end{split}
\eqlabel{h3decompose}
\end{equation}
where we defined 
\begin{equation}
\begin{split}
&h_3=db_2-b_1\wedge d(A)\,,\qquad h_1^\Om=db^\Om-3i\ A\ b^\Om\equiv Db^\Om\,,\\
&h_2=db_1\,,\qquad h_0^\Om=3 i\ b^\Om\,,\\
&h_1^J=db^J-2 b_1\equiv Db^J\,,\qquad h_1^\Phi=db^\Phi-p\ A\equiv Db^\Phi\,. 
\end{split}
\eqlabel{shdef}
\end{equation}

Reducing NSNS 3-form contribution in \eqref{10action} on $T^{1,1}$ results in
\begin{equation}
\begin{split}
&\frac{1}{2\k_{10}^2}\int_{\calm_{10}}\left(-\frac 12 e^{-\phi} H_3^2\right)\star_{10} 1=
-\frac{1}{2\k_5^2}\int_{\calm_5} \biggl\{e^{-4u-\phi}\biggl[(\cosh^2\t\cosh(4\w)-\sinh^2\t)(h_1^J)^2\\
&+(\cosh^2\t\cosh(4\w)+\sinh^2\t)(h_1^\Phi)^2+\cosh^2\t|h_1^\Om|^2-\sinh^2\Re(e^{-2i \theta}(h_1^\Om)^2)
\\
&-2 \cosh^2\t\sinh(4\w) h_1^Jh_1^\Phi -2\sinh(2\t)\left(\sinh(2\w) h_1^J
-\cosh(2\w)h_1^\Phi\right)\Re(i e^{-i \theta}h_1^\Om)
\biggr]\\
&+\frac{1}{2}e^{\ft 83 u-\ft 43 v-\phi}\ h_2^2+\frac 12 e^{\ft {16}{3} u+\ft 43 v-\phi}\ h_3^2+ 
e^{-\ft{20}{3}u-\ft 83 v-\phi} \biggl[\Re(-e^{-2i\theta}\sinh^2\t (h_0^\Om)^2\\
&+2i p  e^{-i \theta}\sinh(2\t)\cosh(2\w)h_0^\Om)
+\cosh^2\t |h_0^\Om|^2+p^2 (\cosh^2\t\cosh(4\w)+\sinh^2\t)
\biggr]
\biggr\} \star 1\,.
\end{split}
\eqlabel{H3red}  
\end{equation}

Similarly, most general $SU(2)\times SU(2)$ symmetric ansatz of RR 3-form field strength $F_3$ 
(solving the Bianchi identity \eqref{flbianchi}) is 
parameterized by a 2-form $c_2$, a one form $c_1$, two real 0-forms $c^J$ and $c^\Phi$, a 
complex 0-form $c^\Omega$ on $\calm_5$ and a constant $q$, 
 \cite{Cassani:2010na}:
\begin{equation}
\begin{split}
F_3=&q\ \Phi\wedge \eta+ d(C_2)-C_0 H_3\,,\\
C_2=&c_2+ c_1\wedge (\eta+A)+c^J J+\Re(c^\Om\Om)+c^\Phi\Phi\,.
\end{split}
\eqlabel{defc2g3}
\end{equation}
The field strength $F_3$ can be decomposed in a basis of left-invariant forms on $T^{1,1}$ \eqref{lif}:
\begin{equation}
\begin{split}
F_3=&g_3+g_2\wedge (\eta+A)+g_1^J\wedge J+\Re\left[g_1^\Om\wedge \Om+g_0^\Om\ \Om\wedge (\eta+A)\right]\\
&+g_1^\Phi\wedge \Phi+(q-C_0 p)\ \Phi\wedge (\eta+A)\,,
\end{split}
\eqlabel{g3decompose}
\end{equation}
where we defined 
\begin{equation}
\begin{split}
&g_3=dc_2-c_1\wedge d(A)-C_0 h_3\,,\qquad g_1^\Om=dc^\Om-3i\ A\ c^\Om-C_0 Db^\Om\equiv Dc^\Om-C_0 Db^\om\,,\\
&g_2=dc_1-C_0db_1\,,\qquad g_0^\Om=3 i\ (c^\Om-C_0 b^\Om)\,,\\
&g_1^J=dc^J-2 c_1-C_0Db^J\equiv Dc^J-C_0Db^J\,,\\
&g_1^\Phi=dc^\Phi-q\ A-C_0 Db^\Phi\equiv Dc^\Phi-C_0Db^\Phi\,. 
\end{split}
\eqlabel{sgdef}
\end{equation}

Reducing RR 3-form contribution in \eqref{10action} on $T^{1,1}$ results in expression 
equivalent to the RHS of \eqref{H3red} with the obvious substitutions:
\begin{equation}
\frac{1}{2\k_{10}^2}\int_{\calm_{10}}\left(-\frac 12 e^{\phi} F_3^2\right)\star_{10} 1=
-\frac{1}{2\k_5^2}\int_{\calm_5}\biggl\{\phi\to -\phi\,, h\to g\,, p\to (q-C_0 p)\biggr\}\,.
\eqlabel{F3red}  
\end{equation}

\subsection{5-form ansatz and  its  reduction reduction}
Because of the self-duality condition \eqref{5self}, special care should be taken 
in dealing with the reduction of the 5-form; furthermore, to reproduce correct type IIB
supergravity equations of motion the 5-form topological term (the second line in \eqref{10action})
must be replaced with \cite{Cassani:2010na}
\begin{equation}
\begin{split}
S_{IIB,top}=&-\frac{1}{8\k_{10}^2}\int_{\calm_5}\biggl[\biggl(B_2\wedge(d(C_2)+2F_3^{fl})-C_2\wedge(d(B_2)
+2H_3^{fl})\biggr)\wedge (d(C_4)+F_5^{fl})\\
&+\frac 12 \bigg(B_2\wedge B_2\wedge d(C_2)\wedge F_3^{fl}+C_2\wedge C_2\wedge d(B_2)\wedge H_3^{fl}\biggr)\biggr]\\
&\equiv -\frac{1}{8\k_{10}^2}\int_{\calm_5}\biggl[L_5\wedge  (d(C_4)+F_5^{fl})+L_{10}\biggr]\,,
\end{split}
\eqlabel{5formtop}
\end{equation} 
where the third line is used to define $L_5$ and $L_{10}$, and 
\begin{equation}
F_3^{fl}=q\ \Phi\wedge \eta\,,\qquad H_3^{fl}=p\ \Phi\wedge \eta\,,\qquad F_5^{fl}=k\ J\wedge J\wedge (\eta+A) \,,
\eqlabel{deffl}
\end{equation}
for a constant $k$. Note that neither $L_5$ nor $L_{10}$  contain 5-form degrees of freedom.
The proper strategy in dealing with the 5-form self-duality condition was developed in \cite{Cassani:2010uw},
which we apply here.

Let's focus first on 5-from degrees of freedom. 5-form Bianchi identity  \eqref{flbianchi} is solved with 
\begin{equation}
F_5=d(C_4)+F_5^{fl}+\frac 12 L_5\,,
\eqlabel{5biachi}
\end{equation}
and the 5-form part of the action \eqref{10action} can be written as 
\begin{equation}
S_{F_5}=-\frac{1}{8\k_{10}^2}\int_{\calm_{10}} \biggl[F_5\wedge \star_{10} F_5+L_5\wedge F_5\biggr]\,.
\eqlabel{5forma}
\end{equation}
As with 3-forms, we can decompose 5-from into the basis of left invariant forms on $T^{1,1}$:
\begin{equation}
\begin{split}
F_5=&f_5+f_4 \wedge (\eta+A)+f_3^J \wedge J+f_2^J\wedge J\wedge (\eta+A)+\Re\left[f_3^\Om\wedge \Om+f_2^\Om\wedge \Om\wedge 
(\eta+A)\right]\\
&+f_3^\Phi\wedge\Phi+f_2^\Phi\wedge\Phi\wedge(\eta+A)+f_1\wedge J\wedge J+f_0\ J\wedge J\wedge (\eta+A)\,,
\end{split}
\eqlabel{5fromdec}
\end{equation}
with 
\begin{equation}
\begin{split}
f_0=k+p c^\Phi-q b^\Phi+3 \Im\left[b^\Om\overline{c^\Om}\right]\equiv k+\frac 12 \ell_0\,,
\end{split}
\eqlabel{f0def}
\end{equation}
\begin{equation}
\begin{split}
f_1&=Da+\frac 12 (q b^\Phi-p c^\Phi) A+\frac 12 \biggl[
b^J Dc^J-b^\Phi Dc^\Phi+\Re\left[b^\Om\overline{Dc^\Om}\right]-b\leftrightarrow c
\biggr]\\&\equiv Da+\frac 12 \ell_1\,,
\end{split}
\eqlabel{f1def}
\end{equation}
\begin{equation}
\begin{split}
f_2^J=d(a_1^J)+\frac 12 \biggl[b^Jd(c_1)-b_1\wedge Dc^J-b\leftrightarrow c\biggr]
\equiv d(a_1^J)+\frac 12 \ell_2^J\,,
\end{split}
\eqlabel{f2jdef}
\end{equation}
\begin{equation}
\begin{split}
f_2^\Om=Da_1^\Om+3 i a_2^\Om+\frac 12 \biggl[b^\Om d(c_1)-b_1\wedge Dc^\Om+3i c^\Om b_2-b\leftrightarrow c\biggr]
\equiv Da_1^\Om+3 i a_2^\Om+\frac 12 \ell_2^\Om\,,
\end{split}
\eqlabel{f2Omdef}
\end{equation}
\begin{equation}
\begin{split}
f_2^\Phi&=d(a_1^\Phi)+\frac 12 (qb_1-p c_1)A+q b_2-p c_2+\frac 12 \biggl[b^\Phi d(c_1)-b_1\wedge Dc^\Phi
-b\leftrightarrow c
\biggr]\\
&\equiv d(a_1^\Phi)+\frac 12 \ell_2^\Phi\,,
\end{split}
\eqlabel{f2Phidef}
\end{equation}
\begin{equation}
\begin{split}
f_3^\Om&=Da_2^\Om-a_1^\Om\wedge d(A)+\frac 12 \biggl[b_2\wedge Dc^\Om+b^\Om(d(c_2)-c_1\wedge d(A))
-b\leftrightarrow c
\biggr]\\
&\equiv Da_2^\Om-a_1^\Om\wedge d(A)+\frac 12 \ell_3^\Om\,,
\end{split}
\eqlabel{f3Omdef}
\end{equation}
\begin{equation}
\begin{split}
f_3^\Phi=&d(a_2^\Phi)-a_1^\Phi\wedge d(A)+\frac 12\biggl[p c_2 \wedge A-q b_2 \wedge A\biggr]
\\&+\frac 12\biggl[b_2\wedge Dc^\Phi+b^\Phi(d(c_2)-c_1\wedge d(A))-b\leftrightarrow c\biggr]
\equiv d(a_2^\Phi)-a_1^\Phi\wedge d(A)+\frac 12 \ell_3^\Phi\,,
\end{split}
\eqlabel{f3Phidef}
\end{equation}
\begin{equation}
\begin{split}
f_3^J&=d(a_2^J)-2 a_3-a_1^J\wedge d(A)+\frac 12 \biggl[b_2 \wedge Dc^J+b^J(d(c_2)-c_1\wedge d(A))-b\leftrightarrow c\biggr]
\\&\equiv d(a_2^J)-2 a_3-a_1^J\wedge d(A)+\frac 12 \ell_3^J\,,
\end{split}
\eqlabel{f3Jdef}
\end{equation}
\begin{equation}
\begin{split}
f_4=d(a_3)+\frac 12 \biggl[b_2 \wedge d(c_1)-b_1\wedge(d(c_2)-c_1\wedge d(A))-b\leftrightarrow c\biggr]
\equiv d(a_3)+\frac 12 \ell_4\,,
\end{split}
\eqlabel{f4def}
\end{equation}
\begin{equation}
\begin{split}
f_5&=f_5^{flux}+d(a_4)-a_3\wedge d(A)+\frac 12 \biggl[b_2\wedge (d(c_2)-c_1\wedge d(A))-b\leftrightarrow c\biggr]
\\&\equiv f_5^{flux}+d(a_4)-a_3\wedge d(A)+\frac 12 \ell_5\,,
\end{split}
\eqlabel{f5def}
\end{equation}
where we defined
\begin{equation}
\begin{split}
Da=&d(a)-2 a_1^J-k A\,,\\
Da_1^\Om=&d(a_1^\Om)-3 i A\wedge a_1^\Om\,,\\
Da_2^\Om=&d(a_2^\Om)-3 i A\wedge a_2^\Om\,.
\end{split}
\eqlabel{defa}
\end{equation}
The last identities  in \eqref{f0def}-\eqref{f5def} are used to define $\{\ell_0,\ell_1,\ell_2^J,\ell_2^\Om,\ell_2^\Phi,
\ell_3^\Om,\ell_3^J,\ell_3^\Phi,\ell_4,\ell_5\}$. 
The form fields $\{a,a_1^J,a_1^\Phi, a_1^\Om,a_2^J,a_2^\Phi, a_2^\Om,a_3, a_4\}$ are degrees of freedom of $C_4$:
\begin{equation}
\begin{split}
&d(C_4)=d(a_4)-a_3 \wedge d(A)+d(a_3) \wedge (\eta+A)+(d(a_2^J)-2 a_3-a_1^J\wedge d(A)) \wedge J
\\&+d(a_1^J)\wedge J\wedge (\eta+A)+\Re\left[(Da_2^\Om-a_1^\Om\wedge d(A))\wedge \Om+(Da_1^\Om+3 i a_2^\Om)\wedge \Om\wedge 
(\eta+A)\right]\\
&+(d(a_2^\Phi)-a_1^\Phi\wedge d(A))\wedge\Phi+d(a_1^\Phi)\wedge\Phi\wedge(\eta+A)+(d(a)-2a_1^J)\wedge J\wedge J\,.
\end{split}
\eqlabel{dc4}
\end{equation} 
Note that given \eqref{dc4}, $d^2 (C_4)=0$.
The self-duality of the 5-form \eqref{5self} relates $\{f_5,f_4,f_3^J,f_3^\Phi,f_3^\Om\}$ to the
remaining 5-form components in \eqref{5fromdec} as follows: 
\begin{equation}
\begin{split}
f_5=&2e^{-\ft{32}{3}u-\ft 83 v}\ \star f_0\,, 
\end{split}
\eqlabel{dualf5}
\end{equation}
\begin{equation}
\begin{split}
f_4=&-2 e^{-8 u}\ \star f_1\,,
\end{split}
\eqlabel{dualf4}
\end{equation}
\begin{equation}
\begin{split}
f_3^J=&e^{-\ft 43 u-\ft 43 v}\ \star\biggl[
(\cosh^2\t\cosh(4\w)-\sinh^2\t) f_2^J-\cosh^2\t\sinh(4\w)f_2^\Phi\\
&-\sinh(2\t)\sinh(2\w)\Re\left(
i e^{-i\theta} f_2^\Om
\right)
\biggr]\,,
\end{split}
\eqlabel{dualf3J}
\end{equation}
\begin{equation}
\begin{split}
f_3^\Phi=&e^{-\ft 43 u-\ft 43 v}\ \star\biggl[
\cosh^2\t\sinh(4\w) f_2^J-(\cosh^2\t\cosh(4\w)+\sinh^2\t)f_2^\Phi\\
&-\sinh(2\t)\cosh(2\w)\Re\left(
i e^{-i\theta} f_2^\Om\right)\biggr]\,,
\end{split}
\eqlabel{dualf3Phi}
\end{equation}
\begin{equation}
\begin{split}
f_3^\Om=&e^{-\ft 43 u-\ft 43 v}\ \star\biggl[
i e^{i\theta}\sinh(2\t)\sinh(2\w) f_2^J-i e^{i\theta}\sinh(2\t)\cosh(2\w)f_2^\Phi
+\cosh^2\t f_2^\Om\\&-\sinh^2\t e^{2i \theta}\overline{f_2^\Om}
\biggr]\,.
\end{split}
\eqlabel{dualf3Om}
\end{equation}
We can not substitute \eqref{dualf5}-\eqref{dualf3Om} directly into  \eqref{5forma}; 
rather, we supplement it with the following  term\footnote{This term is a total derivative on-shell.}:
\begin{equation}
\begin{split}
S'_{F_5}=&\frac{1}{2\k_{10}^2}\int_{\calm_5}\biggl\{
\left(f_5-\frac 12 \ell_5\right) k -\left(f_4-\frac 12 \ell_4\right)\wedge Da\\
&+\left(f_3^J+a_1^J\wedge d(A)-\frac 12 \ell_3^J\right)\wedge d(a_1^J)\\
&+\Re\left[\left(f_3^\Om-Da_2^\Om+d(A)\wedge a_1^\Om-\frac 12 \ell_3^\Om\right)\wedge 
\overline{Da_1^\Om+3i a_2^\Om}\right]\\
&-\biggl(f_3^\Phi+a_1^\Phi\wedge d(A)-\frac 12 \ell_3^\Phi\biggr)\wedge d(a_1^\Phi)
\biggr\}
\wedge \biggl\{\frac 12 J\wedge J\wedge \eta\biggr\}\,.
\end{split}
\eqlabel{sprime}
\end{equation}
In the modified action $S_{F_5}+S'_{F_5}$, the self-duality constraints \eqref{dualf5}-\eqref{dualf3Om} arise 
as equations of motion:
\begin{equation}
\begin{split}
&\frac{\dd}{\dd f_5} \left(S_{F_5}+S'_{F_5}\right)=0\,,\qquad \frac{\dd}{\dd f_4} \left(S_{F_5}+S'_{F_5}\right)=0
\,,\qquad \frac{\dd}{\dd f_3^J} \left(S_{F_5}+S'_{F_5}\right)=0\,,\\
&\frac{\dd}{\dd f_3^\Phi} \left(S_{F_5}+S'_{F_5}\right)=0\,,\qquad \frac{\dd}{\dd \Re[f_3^\Om]} \left(S_{F_5}+S'_{F_5}\right)=0
\,,\qquad \frac{\dd}{\dd \Im[f_3^\Om]} \left(S_{F_5}+S'_{F_5}\right)=0\,.
\end{split}
\eqlabel{eoms5form}
\end{equation} 
The reduced 5-form action is then obtained from imposing the  self-duality constraints \eqref{dualf5}-\eqref{dualf3Om}
in 
\begin{equation}
S_{F_5}^{reduced}=\biggl\{-\frac{1}{8\k_{10}^2}\int_{\calm_10} L_5\wedge F_5+S'_{F_5}\biggr\}\bigg|_{F_5=\star_{10}F_5}
= S_{F_5}^{kinetic}+S_{F_5}^{topological}\,,
\eqlabel{s5red}
\end{equation}
where (up to total derivatives)
\begin{equation}
\begin{split}
&S_{F_5}^{kinetic}=-\frac{1}{2\k_5^2}\int_{\calm_5} \biggl\{
2 e^{-8 u } f_1^2+e^{-\ft 43 u -\ft 43 v }\biggl[\left(\cosh^2\t \cosh(4\w)-\sinh^2\t\right)(f_2^J)^2\\
&+\left(\cosh^2\t\cosh(4\w)+\sinh^2\t\right)
(f_2^\Phi)^2-\sinh^2\t \Re\left(e^{-2 i\theta}(f_2^\Om)^2\right)
+\cosh^2\t |f_2^\Om|^2\\&-2\cosh^2\t\sinh(4\w) f_2^J f_2^\Phi
-2\sinh(2\t)\left(\sinh(2\w)f_2^J-\cosh(2\w)f_2^\Phi\right)\Re\left(i e^{-i\theta}f_2^\Om\right)
\biggr]\\
&+2e^{-\ft{32}{3}u-\ft 83 v} f_0^2
\biggr\} \star 1\,,
\end{split}
\eqlabel{f5kin}
\end{equation}
\begin{equation}
\begin{split}
&S_{F_5}^{topological}=\frac{1}{2\k_5^2}\int_{\calm_5}\biggl\{
\frac i3 \overline{\left(Da_1^\Om+3 ia_2^\Om\right)}\wedge D\left(Da_1^\Om+3 i a_2^\Om\right)
+A\wedge d(a_1^J)\wedge d(a_1^J)\\
&-A\wedge d(a_1^\Phi)\wedge d(a_1^\Phi)-\frac 12 \Re\left[\left(Da_1^\Om+3i a_2^\Om+f_2^\Om\right)
\wedge \overline{\ell_3^\Om}\right]-\frac 12 (d(a_1^J)+f_2^J)\wedge \ell_3^J\\
&+\frac 12 (d(a_1^\Phi)+f_2^\Phi)\wedge \ell_3^\Phi+\frac 12 (Da+f_1)\wedge \ell_4-\frac 12(k+f_0)\wedge \ell_5
\biggr\}\,,
\end{split}
\eqlabel{f5top}
\end{equation}
where we defined
\begin{equation}
D\left(Da_1^\Om+3 i a_2^\Om\right)=d\left(Da_1^\Om+3 i a_2^\Om\right)-3 iA\wedge (Da_1^\Om+3 i a_2^\Om)\,.
\eqlabel{DD}
\end{equation}

Additional contribution to five-dimensional topological couplings comes from $L_{10}$ term in  \eqref{5formtop},
which, up to total derivatives, takes form:
\begin{equation}
\begin{split}
S_{F_5}^{topological,extra}=&\frac{1}{2\k_5^2}\int_{\calm_5} \frac 12 \biggl[p(c_2+c_1\wedge A)-q(b_2+b_1\wedge A)\biggr]
\wedge \biggl[c^\Phi d(b_2+b_1\wedge A)\\
&-b^\Phi d(c_2+c_1\wedge A)\biggr]\,.
\end{split}
\eqlabel{topextra}
\end{equation}

\subsection{CF effective action}

Collecting \eqref{ehred}, \eqref{axidilaton},  \eqref{H3red}, \eqref{F3red}, \eqref{f5kin}, \eqref{f5top}
and \eqref{topextra} we obtain the CF effective action \cite{Cassani:2010na}:
\begin{equation}
S_{eff}=\frac{1}{2\k_2^2} \int_{\calm_5} R \star 1+S_{kin,scal}+S_{kin,vect}+S_{kin,forms}+S_{top}+S_{pot}\,, 
\eqlabel{seff}
\end{equation}
with 
\begin{equation}
\begin{split}
&S_{kin,scal}=-\frac{1}{2\k_5^2}\int_{\calm_5}\biggl\{
\frac{28}{3} du^2+\frac 43 dv^2+\frac 83 dudv +d\t^2+4\cosh^2 \t\ d\w^2
\\&+\sinh^2\t\ (d\theta-3 A)^2+e^{-4u-\phi}\biggl[(\cosh^2\t\cosh(4\w)-\sinh^2\t)(h_1^J)^2\\
&+(\cosh^2\t\cosh(4\w)+\sinh^2\t)(h_1^\Phi)^2+\cosh^2\t|h_1^\Om|^2-\sinh^2\Re(e^{-2i \theta}(h_1^\Om)^2)
\\
&-2 \cosh^2\t\sinh(4\w) h_1^Jh_1^\Phi -2\sinh(2\t)\left(\sinh(2\w) h_1^J
-\cosh(2\w)h_1^\Phi\right)\Re(i e^{-i \theta}h_1^\Om)
\biggr]\\
&+e^{-4u+\phi}\biggl[h\to g\biggr]+\frac 12 d\phi^2+\frac 12 e^{2\phi} dC_0^2+2 e^{-8 u}f_1^2
\biggr\} \star 1\,,
\end{split}
\eqlabel{sks}
\end{equation}
\begin{equation}
\begin{split}
&S_{kin,vect}=-\frac{1}{2\k_5^2}\int_{\calm_5}\biggl\{
\frac 12 e^{\ft 83 u +\ft 83 v}\ (dA)^2+\frac{1}{2}e^{\ft 83 u-\ft 43 v-\phi}\ h_2^2
+\frac{1}{2}e^{\ft 83 u-\ft 43 v+\phi}\ g_2^2\\
&+
e^{-\ft 43 u -\ft 43 v }\biggl[\left(\cosh^2\t \cosh(4\w)-\sinh^2\t\right)(f_2^J)^2
+\left(\cosh^2\t\cosh(4\w)+\sinh^2\t\right)
(f_2^\Phi)^2\\&-\sinh^2\t \Re\left(e^{-2 i\theta}(f_2^\Om)^2\right)
+\cosh^2\t |f_2^\Om|^2-2\cosh^2\t\sinh(4\w) f_2^J f_2^\Phi\\&
-2\sinh(2\t)\left(\sinh(2\w)f_2^J-\cosh(2\w)f_2^\Phi\right)\Re\left(i e^{-i\theta}f_2^\Om\right)
\biggr]
\biggr\} \star 1\,,
\end{split}
\eqlabel{svect}
\end{equation}
\begin{equation}
\begin{split}
&S_{kin,forms}=-\frac{1}{4\k_5^2}\int_{\calm_5} e^{\ft {16}{3} u +\ft 43 v } \left(e^{-\phi} h_3^2+e^\phi g_3^2\right) \star 1\,,
\end{split}
\eqlabel{sforms}
\end{equation}
\begin{equation}
\begin{split}
&S_{top}=\frac{1}{2\k_5^2}\int_{\calm_5}\biggl\{
\frac i3 \overline{\left(Da_1^\Om+3 ia_2^\Om\right)}\wedge D\left(Da_1^\Om+3 i a_2^\Om\right)
+A\wedge d(a_1^J)\wedge d(a_1^J)\\
&-A\wedge d(a_1^\Phi)\wedge d(a_1^\Phi)-\frac 12 \Re\left[\left(Da_1^\Om+3i a_2^\Om+f_2^\Om\right)
\wedge \overline{\ell_3^\Om}\right]-\frac 12 (d(a_1^J)+f_2^J)\wedge \ell_3^J\\
&+\frac 12 (d(a_1^\Phi)+f_2^\Phi)\wedge \ell_3^\Phi+\frac 12 (Da+f_1)\wedge \ell_4-\frac 12(k+f_0)\wedge \ell_5
+\frac 12 \biggl[p(c_2+c_1\wedge A)\\
&-q(b_2+b_1\wedge A)\biggr]
\wedge \biggl[c^\Phi d(b_2+b_1\wedge A)
-b^\Phi d(c_2+c_1\wedge A)\biggr]
\biggr\}\,,
\end{split}
\eqlabel{stop}
\end{equation}
\begin{equation}
\begin{split}
&S_{pot}=\frac{1}{2\k_5^2}\int_{\calm_5}\biggl\{
e^{-\ft 83 u -\ft 23 v}\ R_{T^{1,1}}
-2e^{-\ft{32}{3}u-\ft 83 v} f_0^2-e^{-\ft{20}{3}u-\ft 83 v-\phi} \biggl[\Re(-e^{-2i\theta}\sinh^2\t (h_0^\Om)^2
\\&+2i p  e^{-i \theta}\sinh(2\t)\cosh(2\w)h_0^\Om)
+\cosh^2\t |h_0^\Om|^2+p^2 (\cosh^2\t\cosh(4\w)+\sinh^2\t)
\biggr]\\
&-e^{-\ft{20}{3}u-\ft 83 v+\phi}\biggl[
h\to g\,, p\to (q-p C_0)
\biggr]
\biggr\} \star 1\,.
\end{split}
\eqlabel{spot}
\end{equation}

The equations of motion obtained from  \eqref{seff} are equivalent to type IIB supergravity equations of 
motion \cite{private}. Thus, $SU(2)\times SU(2)$ symmetric 
effective action \eqref{seff} provides consistent truncation of type IIB 
supergravity on resolved warped deformed conifolds with fluxes.

\subsection{Consistent truncations to KS/KT effective actions}

There is a consistent truncation of the $SU(2)\times SU(2)$ symmetric CF action to 
 $SU(2)\times SU(2)\times \zet_2 $ sector describing warped deformed conifold with fluxes 
obtained in \cite{Buchel:2010wp,Buchel:2011cc,Berg:2005pd} with the non-vanishing CF fields identified 
 as 
\begin{equation}
\begin{split}
&e^{-\ft 83 u-\ft 23 v} g_{\mu\nu} dx^{\mu} dx^{\nu}=g_{\mu\nu}^{KS}dx^\mu dx^\nu\,,\qquad k=216 \Om_0^{KS}\,,\qquad q=P^{KS}\,,\qquad \phi=\phi^{KS}\,,\\
&\frac 13 e^v=\Om_1^{KS}\,,\qquad \frac {1}{\sqrt{6}} e^{u-\t/2}=\Om_2^{KS}\,,\qquad \frac {1}{\sqrt{6}} e^{u+\t/2}=\Om_3^{KS}\,,\quad
b^\Phi=-3 \left(h_1^{KS}+h_3^{KS}\right)\,,
\\
&\Im[b^\Om]=3\left(h_3^{KS}-h_1^{KS}\right)\,,\qquad 
 \Re[c^\Om]=6\left(h_2^{KS}-\frac{P^{KS}}{18}\right)\,,
\end{split}
\eqlabel{maptoKS}
\end{equation}
where the superscript $ ^{KS}$ corresponds to the parametrization of fields in  \cite{Buchel:2010wp}.

Further (consistent) restriction to a  $SU(2)\times SU(2)\times U(1)$ symmetric sector of \eqref{maptoKS} with 
\begin{equation}
\begin{split}
&\t=0\,,\qquad \Im[b^\Om]=0\,,\qquad  \Re[c^\Om]=0\,,\qquad \left(b^\Phi-\frac kq\right)=-\frac{K}{2P}\,,\\
&e^v=\Om_1= f_2^{1/2}h^{1/4}\,,\qquad e^u=\Om_2=f_3^{1/2}h^{1/4}\,,\qquad q=P\,,
\end{split}
\eqlabel{furthertrunc}
\end{equation}
leads to the warped conifold with fluxes effective action of \cite{Aharony:2005zr}.

\subsection{Decoupling of linearized fluctuations of CF action around KT action}
Here we characterize decoupled linearized fluctuation sectors about $SU(2)\times SU(2)\times U(1)$ truncation of 
CF effective action:
\begin{equation}
\begin{split}
&S_{KT}=\frac {1}{2\k_5^2}\int_{\calm_5}\biggl\{ R-\frac{28}{3}du^2-\frac 43 dv^2-\frac 83 dudv- e^{-4 u-\phi}
(db^\Phi)^2
-\frac 12 d\phi^2\\
&-2 e^{-\ft{32}{3} u-\ft 83 v} (b^\Phi q-k)^2- e^{-\ft{20}{3}u-\ft 83 v+\phi}q^2
+24 e^{-\ft{14}{3}u-\ft 23 v}
-4 e^{-\ft {20}{3}u+\ft 43 v}
\biggr\} \star 1\,.
\end{split}
\eqlabel{cfkt}
\end{equation} 
Analyzing bilinears of the remaining CF modes about \eqref{cfkt} we find that  there are six decoupled sectors involving:
\nxt $\{\dd C_0,\dd A,\dd c^\Phi,\dd a,\dd a_1^J\}$;
\nxt $\{\dd b_2, \dd c_2, \dd a_1^\Phi, \dd c_1,\dd c^J\}$;
\nxt $\{\dd a_1^\Om, \dd a_2^\Om\}$;
\nxt $\{\Re[\dd b^\Om],\Im[\dd c^\Om]\}$;
\nxt $\{\dd \t,  \Im[\dd b^\Om]\equiv \dd b^\Om_2,\Re[\dd c^\Om]\equiv \dd c^\Om_1\}$;
\nxt $\{ \dd \w, \dd b^J, \dd b_1\}$.\\
Notice that $\dd \theta$ does not couple to quadratic order in KT truncation of CF effective action.

In what follows we focus on the last two fluctuation sets: the chiral symmetry breaking sector, 
\begin{equation}
\begin{split}
&S_{\chi cb}\left[\dd \t,  \dd b^\Om_2,  \dd c^\Om_1\right]=\frac{1}{\k_5^2}\int_{\calm_5}\biggl\{
-\frac 12 (d\dd\t)^2-\frac 12 e^{-4u+\phi} (d\dd c^\Om_1)^2-\frac 12 e^{-4u-\phi} (d\dd b^\Om_2)^2
\\&+2 e^{-4 u-\phi} \dd\t db^\Phi d\dd b^\Om_2+6 e^{-\ft{32}{3}u-\ft 83 v} (b^\Phi q-k) \dd b^\Om_2 \dd c^\Om_1
+6  e^{-\ft{20}{3}u-\ft 83 v+\phi} q \dd \t \dd c^\Om_1\\
&-\frac 92 e^{-\ft{20}{3}u-\ft 83 v} \left(e^{-\phi} (\dd b^\Om_2)^2+e^{\phi} (\dd c^\Om_1)^2\right)
-\frac 12 \biggl(2 e^{-\ft{20}{3}u-\ft 83 v+\phi}q^2 +9 e^{-\ft 83u-\ft 83 v} -12 e^{-\ft{14}{3}u-\ft 23 v}
\\&+2 e^{-4 u-\phi} (db^\Phi)^2
\biggr)(\dd \t)^2
\biggr\}\star 1\,,
\end{split}
\eqlabel{csb}
\end{equation}
and the baryonic branch deformation sector,
\begin{equation}
\begin{split}
&S_{baryonic}\left[\dd \w, \dd b^J\,, \dd b_1\right]=\frac{1}{\k_5^2}\int_{\calm_5}\biggl\{
-\frac 14 e^{\ft 83 u-\ft 43 v-\phi} (d\dd b_1)^2- e^{-4u-\phi}\biggl(\frac 12 (d\dd b^J)^2+2 (\dd b_1)^2\\
&-2 d\dd b^J \dd b_1
-4 \dd \w (d\dd b^J-2 \dd b_1) db^\Phi
\biggr)-2 (d\dd \w)^2
+\biggl(-4 e^{-\ft{20}{3}u-\ft 83 v+\phi}q^2 -16 e^{-\ft {20}{3}u+\ft 43 v} \\
&+24 e^{-\ft{14}{3}u-\ft 23 v}
-4 e^{-4 u-\phi} (db^\Phi)^2 \biggr) (\dd \w)^2
\biggr\}\star 1\,.
\end{split}
\eqlabel{baryonic}
\end{equation}

We explicitly verified that with the 
identifications
\begin{equation}
\dd b^\Om_2=-\frac{1}{2P}\dd k_1\,,\qquad c_1^\Om=\frac{P}{3} \dd k_2\,,\qquad \dd \t =-\frac{\dd f }{f_3}\,,
\eqlabel{chiralfluctuations}
\end{equation} 
the effective action $S_{\chi cb}$ is equivalent to the effective action obtained in   
\cite{Buchel:2010wp}.

Effective action $S_{baryonic}$ is a new result. Remarkably, consistent truncation of the baryonic branch 
deformations around generic $SU(2)\times SU(2)\times U(1)$ states of cascading gauge theory requires 
inclusion of a vector field $\dd b_1$, in addition to the supersymmetric scalar modes $\dd w$ 
and $\dd b^J$ identified in \cite{Butti:2004pk}. We also verified that effective action \eqref{baryonic},
reduced\footnote{As we emphasized earlier, such a reduction is not a consistent truncation.} 
with $\dd b_1=0$, is equivalent to the one discussed in \cite{Buchel:2014hja}.  
Notice that $S_{baryonic}$ is invariant under the $\lambda$-gauge symmetry:
\begin{equation}
\dd b^J\to \dd b^J+2\l\,,\qquad \dd \w\to \dd \w\,,\qquad \dd b_1\to \dd b_1+d\lambda\,,
\eqlabel{gaugeinv}
\end{equation}   
for an arbitrary 0-form $\lambda$ on $\calm_{5}$. This gauge symmetry is simply a restriction
of general $\lambda$-gauge transformations discussed in \cite{Cassani:2010na} to linearized 
(decoupled) fluctuations $\{\dd \w, \dd b^J\,, \dd b_1\}$ about  $SU(2)\times SU(2)\times U(1)$ 
states of cascading gauge theory. Gauge symmetry \eqref{gaugeinv} can be used to completely 
eliminate $\dd b^J$ fluctuations.

\section{Baryonic branch in cascading gauge theory plasma}\label{section3}
As an application of the effective action \eqref{baryonic}, we study stability of 
 the baryonic branch fluctuations in cascading gauge theory plasma \cite{Aharony:2007vg}.
We  focus on geometries dual to thermal states
of cascading plasma, and study  the spectrum of the baryonic branch quasinormal modes of Klebanov-Tseytlin
black hole \cite{Aharony:2007vg,Buchel:2009bh}.  We show that these modes remain massive for all 
accessible temperatures, \ie for $T\ge T_{u}$. 

First, we rewrite effective action \eqref{baryonic} using the KS background metric 
(see \eqref{maptoKS}):
\begin{equation}
g_{\mu\nu}\to g_{\mu\nu} \Om^{-2}\,,\qquad \Om=e^{-\ft 43 u-\ft 13 v}\,.
\eqlabel{rescaleg}
\end{equation}
As a result of a Weyl rescaling \eqref{rescaleg}, 
\begin{equation}
\star 1\to \Om^{-5}\ \star 1\,,\qquad A_{(p)}B_{(p)} \to \Om^{2p}\ A_{(p)}B_{(p)} \,,
\eqlabel{weyl}
\end{equation}
for any $p$-forms $ A_{(p)}$ and $ B_{(p)}$ on $\calm_5$. Thus, \eqref{baryonic}
is modified to 
\begin{equation}
\begin{split}
&\hat{S}_{baryonic}\left[\dd \w, \dd b^J\,, \dd b_1\right]=\frac{1}{\k_5^2}\int_{\calm_5}\biggl\{
-\frac 14 e^{4 u -v-\phi} (d\dd b_1)^2-2 e^{4 u+v} (d\ \dd \w)^2\\
&+e^{v-\phi}\biggl(
-2 (\dd b_1)^2-8 \dd \w \dd b_1 db^\Phi-4 (\dd \w)^2 (db^\Phi)^2+2 d\dd b^J \dd b_1
+4\dd \w d\dd b^J db^\Phi\\&-\frac 12 (d\ \dd b^J)^2
\biggr)
+\biggl( -4 e^{-v+\phi} q^2+24 e^{2u+v}-16 e^{3v}\biggr) (\dd \w)^2
\biggr\}\star 1\,.
\end{split}
\eqlabel{flfinal}
\end{equation}
 
 The background geometry  dual to the deconfined 
homogeneous and isotropic phase of the cascading plasma is given by 
\begin{equation}
\begin{split}
&ds_5^2=h^{-1/2}(1-f_1^2)^{-1/2}\left(-f_1^2\ dt^2+dx_1^2+dx_2^2+dx_3^2\right)+\frac 19 h^{1/2}f_2\
\frac{dr^2}{\tf_2^2} \,,\\
&u=\ln \left(f_3^{1/2} h^{1/4}\right)\,,\qquad v=\ln \left(f_2^{1/2} h^{1/4}\right)\,,\qquad db^\Phi=-\frac{1}{2P}dK\,,\qquad 
q=P\,,
\end{split}
\eqlabel{background}
\end{equation} 
with $\{f_1,\tf_2,K,h,f_2,f_3,g_s\equiv e^\phi\}$ being functions of $r$ only.
We focus on modes at the threshold of instability, thus, without loss of generality we 
assume\footnote{Here, we use the gauge symmetry \eqref{gaugeinv} to eliminate $\dd b^J$ 
and assume propagation of quasinormal modes along $x_1$ direction.} 
\begin{equation}
\begin{split}
&\dd b^J=0\,,\qquad \dd w=-\frac 12 e^{i k x_1}\ \calz\,,\\
&\dd b_{1,x_1}=i k e^{i k x_1} \calb_{x_1}\,,\qquad 
\dd b_{1,r}=e^{i k x_1} \calb_r\,,\qquad \dd b_{1,t}= \dd b_{1,x_2}=\dd b_{1,x_3}=0\,,
\end{split}
\eqlabel{flucw}
\end{equation}
where  $\{\calz,\calb_{x_1},\calb_{r}\}$ are functions of the radial coordinate only, satisfying the following 
equations of motion (obtained from \eqref{flfinal})
\begin{equation}
\begin{split}
0=&k^2 f_1^2\ \calz-\frac{9 \tilde{f}_2^2 f_1^2}{h f_2 (1-f_1^2)^{1/2}}\
\calz''
-\frac{9\tilde{f}_2 f_1}{f_2 f_3 h (1-f_1^2)^{3/2}} \biggl(\tilde{f}_2 f_3 f_1' f_1^2-2 \tilde{f}_2 f_3' f_1^3
\\&-f_3 \tilde{f}_2' f_1^3
+\tilde{f}_2 f_3 f_1'+2 \tilde{f}_2 f_3' f_1
+f_3 \tilde{f}_2' f_1)\biggr)\  
 \calz'
-\frac{f_1^2}{2h^2 f_3^2 g_s P^2f_2 (1-f_1^2)^{1/2}} \biggl(\\
&24 h f_3 g_sP^2 f_2-16 h g_s P^2f_2^2 -4 g_s^2 P^4
-9 \tilde{f}_2^2 (K')^2\biggr)\ \calz+\frac{18\tilde{f}_2^2f_1^2 K'}{g_sf_2 f_3^2 (1-f_1^2)^{1/2} P h^2} \calb_r\,,
\end{split}
\eqlabel{zequa}
\end{equation}
\begin{equation}
\begin{split}
0=&\calb_{x_1}''-\frac{1}{\tilde{f}_2 f_3 f_1 f_2 g_s (1-f_1^2)}\biggl(2 \tilde{f}_2 f_2 g_s f_1^3 f_3'+f_3 f_2 g_s f_1^3 \tilde{f}_2'
-2 \tilde{f}_2 f_2 g_s f_1 f_3'
-f_3 \tilde{f}_2 f_1' f_2 g_s\\
&-f_3 f_2 g_s f_1 \tilde{f}_2'-f_3 \tilde{f}_2 g_s' f_2 f_1^3
+f_3 \tilde{f}_2 f_2' g_s f_1+f_3 \tilde{f}_2 g_s' f_2 f_1
-f_3 \tilde{f}_2 f_2' g_s f_1^3\biggr) \calb_{x_1}'\\
&-\frac{8f_2^2}{9f_3^2 \tilde{f}_2^2} \calb_{x_1}
-\calb_r'+\frac{1}{f_3 \tilde{f}_2 f_1 f_2 g_s (1-f_1^2)}\biggl(2 \tilde{f}_2 f_2 g_s f_1^3 f_3'+f_3 f_2 g_s f_1^3 \tilde{f}_2'
-2 \tilde{f}_2 f_2 g_s f_1 f_3'\\
&-f_3 \tilde{f}_2 f_1' f_2 g_s-f_3 f_2 g_s f_1 \tilde{f}_2'
-f_3 \tilde{f}_2 g_s' f_2 f_1^3+f_3 \tilde{f}_2 f_2' g_s f_1+f_3 \tilde{f}_2 g_s' f_2 f_1
\\&-f_3 \tilde{f}_2 f_2' g_s f_1^3\biggr) \calb_r\,,
\end{split}
\eqlabel{bxequa}
\end{equation}
\begin{equation}
\begin{split}
0=&\frac{hf_3^2 k^2 (1-f_1^2)^{1/2}}{f_2}\calb_{x_1}'-\frac{hf_3^2f_1^2k^2(1-f_1^2)^{1/2}+8 f_1^2 f_2}{f_1^2 f_2} \calb_r-\frac{4K'}{P}\calz\,.
\end{split}
\eqlabel{brequa}
\end{equation}
Notice that equation \eqref{brequa} can be used to algebraically eliminate $\calb_r$ from equations \eqref{zequa}
and \eqref{bxequa}.

To make use of the results in \cite{Aharony:2007vg,Buchel:2009bh} we use a radial coordinate $x$ 
as 
\begin{equation}
x\equiv 1-f_1(r) \,.
\eqlabel{defx}
\end{equation}
 The physical fluctuations described by \eqref{zequa}-\eqref{brequa}
must be regular at the horizon of the KT BH, 
and be normalizable at the asymptotic $x\to 0_+$ boundary.
Introducing 
\begin{equation}
\kk=\frac{k}{2\pi T}\,,
\eqlabel{wq}
\end{equation} 
and using the asymptotic expansion for the KT BH developed in 
\cite{Aharony:2007vg}\footnote{As explained in \cite{Aharony:2007vg} we can set in numerical analysis $a_0=1$.}, 
the normalizability condition 
for $\{\calz\,, \calb_{x_1}\}$
at the $x\to 0_+$ boundary translates into the following asymptotic solution 
\begin{equation}
\begin{split}
&{\calz}=z_1 x^{1/2}+\frac{\pi^2 T^2 \kk^2 z_1}{4\sqrt{2}}(2k_s+9-\ln x )x 
+\calo(x^{3/2}\ln^2 x)\,,
\end{split}
\eqlabel{uvz}
\end{equation}
\begin{equation}
\begin{split}
&{\calb_{x_1}}=x\left(b_{2,0}+\frac{\pi^2 T^2\kk^2 z_1\sqrt{2}\ln x}{1152}(12 k_s+94-3\ln x)\right)+\calo(x^{3/2}\ln^3 x)\,,
\end{split}
\eqlabel{uvc}
\end{equation}
where we presented the expansions only to leading order in the normalizable 
UV coefficients
\begin{equation}
\biggl\{z_{1}\,, b_{2,0}\biggr\}\,.
\eqlabel{normalizableuv} 
\end{equation}
The independent UV normalizable coefficients \eqref{normalizableuv} imply that the 
baryonic branch deformation in cascading plasma is associated with the development of the 
expectation values of operators of dimension-2 and dimension-4.

Since the equations of motion \eqref{zequa}-\eqref{brequa} are homogeneous,
without the loss of generality we can set $\calz(1)=1$. The IR, \ie as $y\equiv (1-x)
\to 0_+$, 
asymptotic expansion then takes form
\begin{equation}
{\calz}=1+\calo(y^2)\,,\qquad {\calb_{x_1}}=b_{0}^h+\calo(y^2)\,,
\eqlabel{irasymptotic}
\end{equation}
where we presented the expansions only to leading order in the normalizable 
IR coefficient
\begin{equation}
\biggl\{b_{0}^h\biggr\}\,.
\eqlabel{normalizableir} 
\end{equation}

\begin{figure}[t]
\begin{center}
\psfrag{l}[BR][T]{{$\ln \frac{T}{\Lambda}$}}
\psfrag{t}{{$ \frac{T}{\Lambda}$}}
\psfrag{q}{{$\kk^2$}}
\includegraphics[width=3.0in]{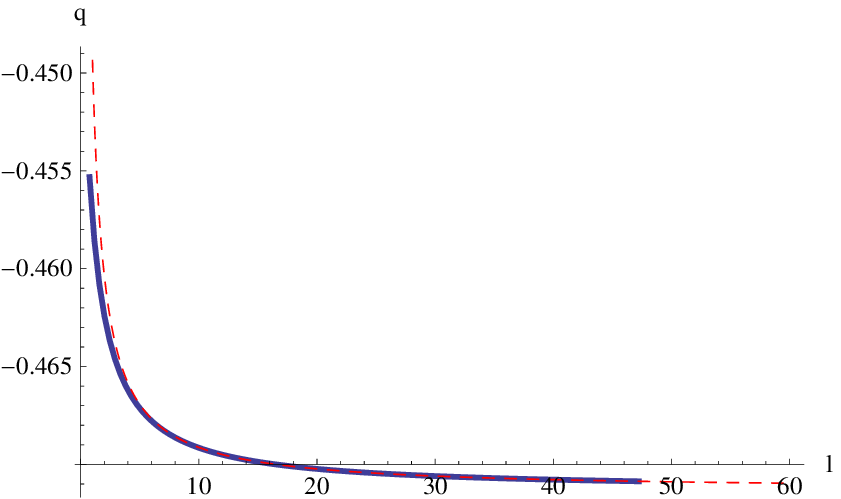}
\includegraphics[width=3.0in]{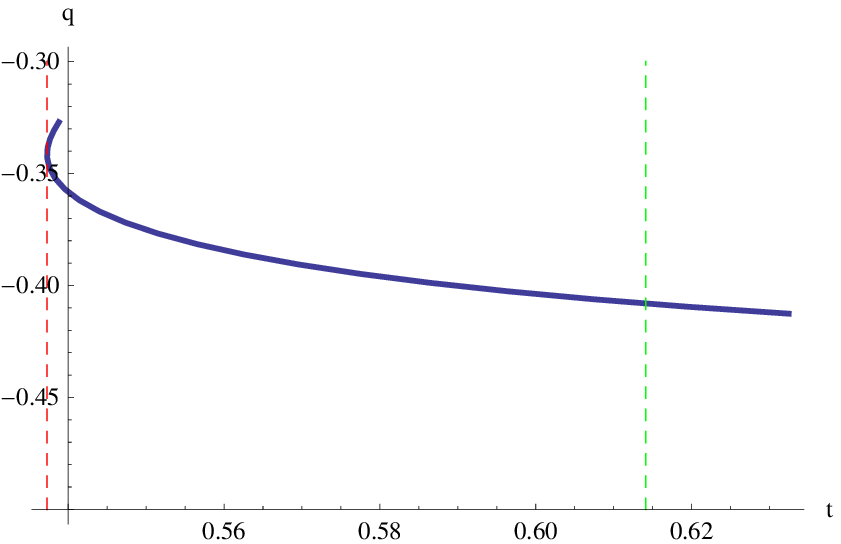}
\end{center}
  \caption{(Colour online)  {\bf Left panel:}  Dispersion relation of the baryonic branch quasinormal modes 
of the Klebanov-Tseytlin black hole as a function of $\ln \frac{T}{\Lambda}$
at high temperature. The solid blue line represents the dispersion relation of the baryonic branch
fluctuations.
The red dashed line is a fit \eqref{fitred} to the data.  {\bf Right panel:} 
Dispersion relation at low temperatures.  The vertical dashed green and red lines 
indicate $T=T_c$ (the confinement/deconfinement temperature) and $T=T_u$ (the hydrodynamic instability 
temperature) correspondingly.
 } \label{figure1}
\end{figure}

The results of the analysis of the dispersion relation of the baryonic branch 
quasinormal modes are presented in Figure~\ref{figure1}.
In principle, we expect discrete branches of the quasinormal modes
distinguished by the number of nodes in radial profiles 
$\{\calz\,,{\calb_{x_1}}\}$. In what follows we consider 
only the lowest quasinormal mode, which has monotonic radial profiles. 
We find that over all range of temperatures, the fluctuations (solid blue line) 
have $\kk^2<0$  --- as a result,
they are massive. The red dashed line 
\begin{equation}
\kk^2\bigg|_{\rm{red,dashed}}\ =\ -0.47(1)+0.02(2)\ \ln^{-1}\frac{T}{\Lambda}
+\calo\left(\ln^{-2}\frac{T}{\Lambda}\right) \,,
\eqlabel{fitred}
\end{equation}
represents the best fit to (the high-temperature tail of)  the data. 
Notice that in the limit $T\gg \Lambda$ the cascading theory approaches a 
conformal theory with temperature being the only relevant scale, thus, 
in agreement with \eqref{fitred}, $\kk^2$
must approach a constant in this limit.

\section*{Acknowledgments} I would like to 
thank Ofer Aharony for useful discussion. 
I am particularly grateful to Davide Cassani and Anton Faedo for comments on the 
first draft of this paper.
I would like to thank Weizmann Institute of Science and APCTP
 for hospitality during 
the completion of this work. 
Research at Perimeter
Institute is supported by the Government of Canada through Industry
Canada and by the Province of Ontario through the Ministry of Research
\& Innovation. I gratefully acknowledge support from NSERC
Discovery grant. 

\appendix

\section{Conventions}\label{apA}
A differential $p$-form $A_{(p)}$ in ten dimensions is defined as 
\begin{equation}
A_{(p)}=\frac {1}{p!}\ A_{(p)\, I_1 \cdots I_p }\ \uE^{I_1}\wedge \cdots\wedge \uE^{I_p}\,,
\eqlabel{defform}
\end{equation}
where $A_{(p)\, I_1\cdots I_p }$ are form components in orthonormal ten-dimensional vielbein  $\{\uE^I\}$ basis. 
A Hodge dual is defined according to 
\begin{equation}
\star_{10}\ \uE^{I_1}\wedge\cdots\wedge \uE^{I_p}=\frac{1}{(10-p)!}\e^{I_1\cdots I_p}_{\qquad I_{p+1}\cdots I_{10}}\ 
\uE^{I_{p+1}}\wedge \cdots\wedge \uE^{I_{10}}\,,
\eqlabel{hodge10}
\end{equation}
with 
\begin{equation}
\e_{1\cdots 10}=+1\,,\qquad \e^{1\cdots 10}=-1\,.
\eqlabel{defep10}
\end{equation}
Similarly, a differential $p$-form $A_{(p)}$ in five dimensions is defined as 
\begin{equation}
A_{(p)}=\frac {1}{p!}\ A_{(p)\, i_1 \cdots i_p }\ E^{i_1}\wedge \cdots\wedge E^{i_p}\,,
\eqlabel{defform5}
\end{equation}
where $A_{(p)\, i_1\cdots i_p }$ are form components in orthonormal five-dimensional vielbein  $\{E^i\}$ basis. 
A Hodge dual is defined according to 
\begin{equation}
\star\ E^{i_1}\wedge\cdots\wedge E^{i_p}=\frac{1}{(5-p)!}\e^{i_1\cdots i_p}_{\qquad i_{p+1}\cdots i_{5}}\ 
E^{i_{p+1}}\wedge \cdots\wedge E^{i_{10}}\,,
\eqlabel{hodge5}
\end{equation}
with 
\begin{equation}
\e_{1\cdots 5}=+1\,,\qquad \e^{1\cdots 5}=-1\,.
\eqlabel{defep5}
\end{equation}
Given two $p$-forms $A_{(p)}$ and $B_{(p)}$ we have
\begin{equation}
\begin{split}
&A_{(p)}\wedge \star_{10} B_{(p)}= \left[ \frac{1}{p!}\ A_{(p)I_1\cdots I_p} B_{(p)}^{I_1\cdots I_p}\right]\star_{10} 1\equiv 
\left[A_{(p)}B_{(p)}\right]\star_{10} 1  \,,\\
&A_{(p)}\wedge \star B_{(p)}= \left[\frac{1}{p!}\ A_{(p)i_1\cdots i_p} B_{(p)}^{i_1\cdots i_p}\right]\star 1
\equiv \left[A_{(p)}B_{(p)}\right]\star 1 \,,
\end{split}
\eqlabel{wedgeprod}
\end{equation}  
in ten and five dimensions correspondingly.

\end{document}